\documentclass[sort&compress,preprint,review]{elsarticle}
\usepackage{lineno,hyperref}
\usepackage{textcomp}
\usepackage{setspace}
\usepackage{color}
\journal{Advances in Colloid and Interface Science}

\begin{document}

\begin{frontmatter}

\title{
Surface Nanobubbles: Theory, Simulation, and Experiment. A Review
%\tnoteref{mytitlenote}
}


\author[panos]{Panagiotis E. Theodorakis\corref{mycorrespondingauthor1}}
\cortext[mycorrespondingauthor1]{panos@ifpan.edu.pl}

\author[zhizhao]{Zhizhao Che\corref{mycorrespondingauthor2}}
\cortext[mycorrespondingauthor2]{chezhizhao@tju.edu.cn}

\address[panos]{Institute of Physics, Polish Academy of Sciences, Al.\ Lotnik\'ow 32/46, 02-668 Warsaw, Poland}
\address[zhizhao]{State Key Laboratory of Engines, Tianjin University, Tianjin, 300072, China}

\begin{abstract}
Surface nanobubbles (NBs) are stable gaseous phases in liquids that form
at the interface with solid substrates. They have been particularly intriguing for their high stability
that contradicts theoretical expectations and their \textit{potential} relevance for many technological
applications. Here, we present the current
state of the art in this research area by discussing and contrasting main results obtained from
theory, simulation and experiment, and presenting their limitations. We also provide future
perspectives anticipating that this review will stimulate
further studies in the research area of surface NBs.
\end{abstract}

\begin{keyword}
Surface Nanobubbles \sep Experiment \sep Theory \sep Simulation \sep Contact-line Pinning \sep Oversaturation
\end{keyword}

\end{frontmatter}

%\linenumbers
%\begin{figure}[b]
%    \centering
%    \includegraphics[width=\columnwidth]{figure00-toc.pdf}
%    \label{fig:fig-toc}
%\end{figure}

%Figure captions appear in the end and figure files are sent separately
% \begin{figure}
% \caption{Each figure legend should begin with a brief title for
% the whole figure and continue with a short description of each
% panel and the symbols used. For contributions with methods
% sections, legends should not contain any details of methods, or
% exceed 100 words (fewer than 500 words in total for the whole
% paper). In contributions without methods sections, legends should
% be fewer than 300 words (800 words or fewer in total for the whole
% paper).}
% \end{figure}

\section{Introduction}\label{sec:sec1-intro}
\begin{figure}
    \centering
    \includegraphics[width=\columnwidth]{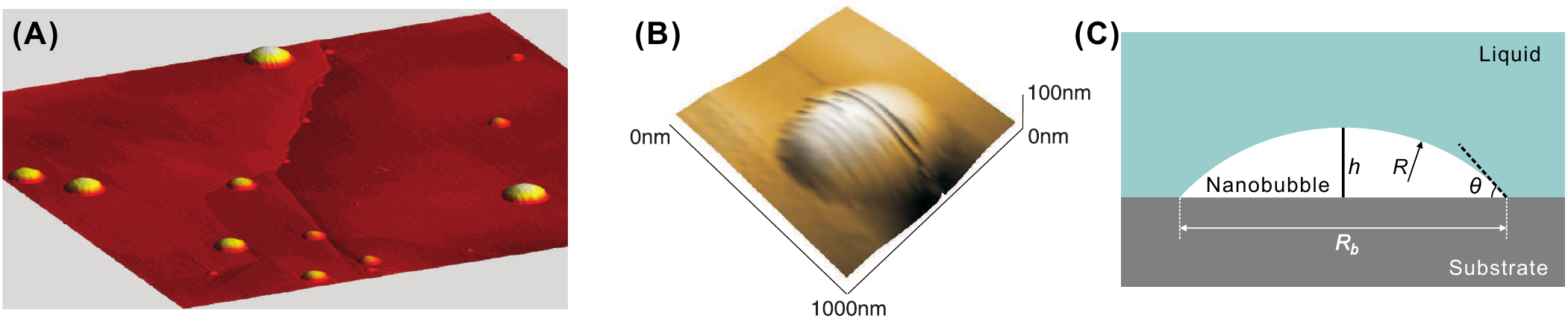}
    \caption{Surface nanobubbles. (A) Nanobubbles on an HOPG surface immersed in water. Reprinted with permission from Ref.~\cite{Borkent2010}. Copyright 2010 American Chemical Society ; (B) Three-dimensional topological image of a single nanobubble on HOPG substrate at room temperature. Reprinted from Ref.~\cite{Seddon2011a}; (C) Schematic illustration of NB shape and parameters characterising its structure.}
    \label{fig:fig1}
\end{figure}

Surface nanobubbles (NBs) are gaseous domains at the interface between a
liquid medium (\textit{e.g.}\ water) and a solid substrate,
which is usually hydrophobic (see Fig.~\ref{fig:fig1})  \cite{Lohse2015,Peng2015a,Alheshibri2016,Craig2011}.
The existence of NBs was speculated about 25 years ago by Parker
\textit{et al.}\ as they attempted to estimate the forces between two neutral hydrophobic surfaces
immersed in water \cite{Parker1994}. In this case, discontinuities in the force
measurements were attributed to microscopic bubbles or cavities between the surfaces.
By carrying out  Atomic Force Microscopy (AFM) measurements, Carambassis \textit{et al.}\ later suggested
that a long-range attraction between the immersed surfaces was
due to long-lived submicron bubbles \cite{Carambassis1998}.
Subsequent investigations have drawn similar conclusions in the case of hydrophobised silica
surfaces providing initial evidence of NBs existence \cite{Gong1999,Yakubov2000}.

Despite these early studies, the existence of stable surface NBs was established by Lou \textit{et al.},
who provided the first AFM image of hour-long stable NBs on a mica surface immersed in water \cite{Lou2000}.
The acquisition of AFM images was possible due to the apparently slower relaxation of
the NBs in comparison with the motion of the AFM tip,
which could sense a repulsive force during
the interaction with the NBs. In the same year, images of surface NBs
on silicon wafer hydrophobised with octadecyltrichlorosilane (OTS)
were obtained by Ishida \textit{et al.}, providing further evidence for the existence of surface NBs \cite{Ishida2000}.
These early studies indicated that the shape of NBs resembled roughly a spherical cap with height
ranging from one to tens of nanometres and
base radius from hundreds of nanometres to several microns.
Moreover, the apparent contact angle
(measured on the gas side, see Fig.~\ref{fig:fig1}) of the NBs was much smaller
than the macroscopically expected values \cite{Ishida2000}.
Further experiments suggested that the contact angle
be also a function of NB size  \cite{Lou2000,Lou2002,Tyrrell2001,Yang2003,Simonsen2004},
while their density be temperature-dependent \cite{Zhang2004}.
Work by Tyrrell \textit{et al.}\ by means of AFM experiments indicated that the
radius of curvature of the NBs was of the order of 100~nm, while the height
was 20--30~nm \cite{Tyrrell2001}.
This particular morphology of surface NBs was believed to
be the reason for their unexpected high stability \cite{Ishida2000,Tyrrell2001}
despite the theoretically predicted
high Laplace pressure ($\Delta p = 2 \gamma / R$, where $\gamma$ is the
surface tension and $R$ the bubble radius, see Fig.~\ref{fig:fig1}) \cite{Epstein1950},
which would suggest their fast dissolution in a time scale of a few
microseconds \cite{Simonsen2004,Ball2003}.
Yet, the stability of surface NBs has been confirmed on different substrates, for example,
in the case of Highly Oriented Pyrolytic Graphite (HOPG) substrates \cite{Zhang2008a, Berkelaar2013}
and ultraflat gold surfaces  \cite{Holmberg2003}, while unstable
surface NBs have been reported in the case of HOPG substrates immersed in alcohol \cite{Zhang2005,Hampton2008}.

The existence of surface NBs has been further established by different experiments,
such as Attenuated Total Reflectance Infrared Spectroscopy  \cite{Zhang2007,Zhang2008b},
Quartz Crystal Microbalance \cite{Seo2007,Zhang2008}, Surface Plasmon
Resonance \cite{Martinez2007, Zhang2007,Zhang2008b},
and Neutron Reflectometry \cite{Steitz2003}. In the case of single NBs, the evidence
stems from Scanning Electron Microscopy (SEM) imaging \cite{Switkes2004}, Scanning Transmission
Soft X-ray Microscopy \cite{Zhang2013}, and Single Photon Counting combined with
Fluorescence Lifetime Imaging Microscopy \cite{Hain2016}. Direct observation
with Interference-Enhanced Reflection Microscopy \cite{Karpitschka2012}
and with Total-Internal-Reflection-Fluorescence Microscopy (TIRF)
has been also recently provided in the literature \cite{Chan2012,Chan2015,Chan2015a}.

%[!!!!too many references in this paragraph!!!!] Owing to their high stability and spontaneous formation, surface NBs are important in the context of many modern applications where the role of surface interactions is important \cite{Christenson2001,Attard2003,Hampton2010,Parker1994,Stevens2005,Mastropietro2012}, such as flotation \cite{Scheludko1976} and separation processes \cite{Mahnke1999,Mishchuk2002,Mishchuk2006,Zhang2011a,Calgaroto2014}. NBs have been also found to play a significant role in the case of nanocomposite foams \cite{Chen2013, Liu2015a}, nanomaterial engineering  \cite{Hui2009,Huang2009,Berkelaar2013,Gao2014,Feuillebois2009,Lee2014a,Ismagilov2002,Paxton2004,Guix2014,Manjare2012,Solovev2009,Wilson2012,Lee2014,Guillemot2012,Wang2017}, catalysis and electrolysis \cite{Fujishima1972,Luo2014,Zakzeski2010,Steinfeld2005,Westerheide1961,Ammam2012,Ahn2013,Hammadi2013,Suslick1999,Vogt2005,Zhang2012a}, plasmonic and vapour NBs \cite{Halas2011,Lukianova-Hleb2010,Adleman2009,Fang2013,Carlson2012,Baffou2014,Boulais2012,Cavicchi2013,Polman2013,Thomann2011}, and transport in nanofluidic devices as well as boiling heat transfer by serving as nucleation sites \cite{Neto2005,Bocquet2010,Hyvaluoma2011,Rothstein2010,Lee2014a,Paxton2004}.
Owing to their high stability and spontaneous formation, surface NBs are important in the context of many modern
applications where the role of surface interactions becomes
relevant \cite{Hampton2010,Mastropietro2012}, such as separation
processes \cite{Mishchuk2002}. Surface NBs have been also found to play a significant role in the case of flotation \cite{Calgaroto2014}, drag reduction \cite{Lohse2018,Kumagai2015}, nanocomposite foams
\cite{Chen2013, Liu2015a}, nanomaterial engineering  \cite{Huang2009}, catalysis and electrolysis \cite{Hammadi2013},
plasmonic and vapour NBs \cite{Lukianova-Hleb2010}, and transport in nanofluidic devices as well as boiling heat transfer
by serving as nucleation sites \cite{Zou2018langmuir}.
The fabrication of nanoporous polypyrrole (PPy) films
is a specific example, where stable NBs can be used as a template
to create nanoporous films \cite{Hui2009}. In this case,
hydrogen surface NBs electrochemically form on a bare HOPG substrate.
Then, electropolymerisation of pyrrole takes place around the surface
NBs. In the end, PPy nanoporous films are obtained by removing the
NBs.

The subsequent sections of this review article are organised as follows: In Sec.~\ref{sec:sec2},
we present a range of different theoretical, simulation and experimental methodologies,
which are used to study surface NBs discussing their limitations and contrasting
main results. In Sec.~\ref{sec:sec3-morphology}, we discuss the main
morphological and mechanical characteristics of surface NBs.
In Sec.~\ref{sec:sec4-formation}, we provide a review of the main
experimental methods to generate surface NBs. In Sec.~\ref{sec:sec-stability},  we
discuss the main hypothesis that supports the intriguing high
NBs' stability, \textit{i.e.} contact-line pinning. Finally, we
present our perspectives in the research area of surface NBs
in Sec.~\ref{sec:sec6-perspectives}.

\section{Methods to study NBs and main results}\label{sec:sec2}
This part of the review describes different methodologies for studying NBs and main results obtained by
theoretical  (Section \ref{sec:sec2-1-theory}), molecular dynamics simulation
(Section \ref{sec:sec2-2-simulation}), and experimental studies (Section \ref{sec:sec2-3-experiment}).
We contrast these results and discuss the limitations of the various methods.

\subsection{Theory}\label{sec:sec2-1-theory}
The rate of dissolution by diffusion of a gas (bulk)
bubble without translational motion in an undersaturated
solution as well as the rate of bubble growth in
an oversaturated solution may be described by the Epstein--Plesset theory \cite{Epstein1950}.
While both solutions (either for the growth or the dissolution process) are similar, the rate of bubble growth is inversely proportional
to the radius of the bubble and the gas diffusivity. %The theory also predicts that the surface tension is inversely proportional to the rate of bubble growth.
Considering estimates
for bubble size in the range 10--100~nm, an extension of the Epstein--Plesset theory
would suggest that the bubbles have short lifetimes \cite{Ljunggren1997}.
%and the bridging bubble mechanism could not be relevant .
Aiming at explaining the high stability of surface NBs,
many theories have been proposed to fill the gap in
our understanding, such as the
contamination at the gas--water interface that hinders the gas exchange
and reduces the surface tension \cite{Ducker2009},
the dynamic equilibrium \cite{Brenner2008} that postulates
that the NBs' stability is due to a balance between gas outflux and
influx, and the pinning of the contact line \cite{Brenner2008,Weijs2013,Zhang2013,Liu2014a,Liu2014b,Lohse2015c}.
%in agreement with results obtained in the case of electrolytically generated NBs  (Fig.~\ref{fig:fig-theory}A) \cite{Yang2009a}.
Weijs and Lohse \cite{Weijs2013},
have initially considered the NB as a source of gas oversaturation, which
dissipates by diffusion as the liquid equilibrates to saturation. Later,
Lohse and Zhang \cite{Lohse2015c} and Chan \textit{et al.}\ \cite{Chan2015a}
have assumed only gas transport in the vicinity of the NB. The latter
two approaches have been recently coupled into a theoretical model, which
can explain the stability and dynamics of surface NBs in undersaturated
environments \cite{Tan2019}.
The framework of the model of Lohse and Zhang \cite{Lohse2015c}
has been recently exploited to describe the coarsening process of
competitively growing NBs that interact diffusively \cite{Zhu2018}.
They found that the coarsening slows down with advancing time
and increasing bubble distance.

Contamination theory has attempted to describe the unique properties of surface NBs
(\textit{i.e.}\ the low contact angle as measured from the gas side, see Fig.~\ref{fig:fig1}, and the high stability) by suggesting that a film of
contaminants would form at the liquid--gas interface, which
may be responsible for a reduction of the surface tension that could eventually lead to the dissociation
of bubbles \cite{Ducker2009}.
In turn, this would allow for a low gas-side contact angle of the NBs (Fig.~\ref{fig:fig1}) and a higher diffusion of
gas outside NBs, which would lead to a higher stability.
This hypothesis has been tested quantitatively
by Das \textit{et al.}\ \cite{Das2010}, who estimated the dependence of the contact angle and
the Laplace pressure on the fraction of impurity coverage at the liquid--gas interface.
According to this estimate, the increase in the fractional coverage of the impurities leads to the
reduction of both the contact angle and the Laplace pressure. An extension of this model can
account for further nonidealities at the liquid--gas interface \cite{Das2011}. Such nonidealities
can predict a significantly lower contact angle and Laplace pressure even at a smaller impurity coverage,
which highlights further the role of impurities in the stability of NBs. Another extension
refers to the consideration of surfactants as `impurities' by Zhang \textit{et al.}\ \cite{Zhang2012b}.
In this case, water-soluble surfactants appeared to have little effect on the stability of NBs, whereas
water-insoluble surfactants dramatically reduced their formation and stability.
However, Zhang \textit{et al.}\  suggest
that NBs stability could not be attributed to a layer of organic contaminants
originating from the HOPG or other surface-active materials,
for example, as a result of the standard solvent exchange method for generating NBs \cite{Zhang2012b}.

\begin{figure}
    \centering
    \includegraphics[width=0.8\columnwidth]{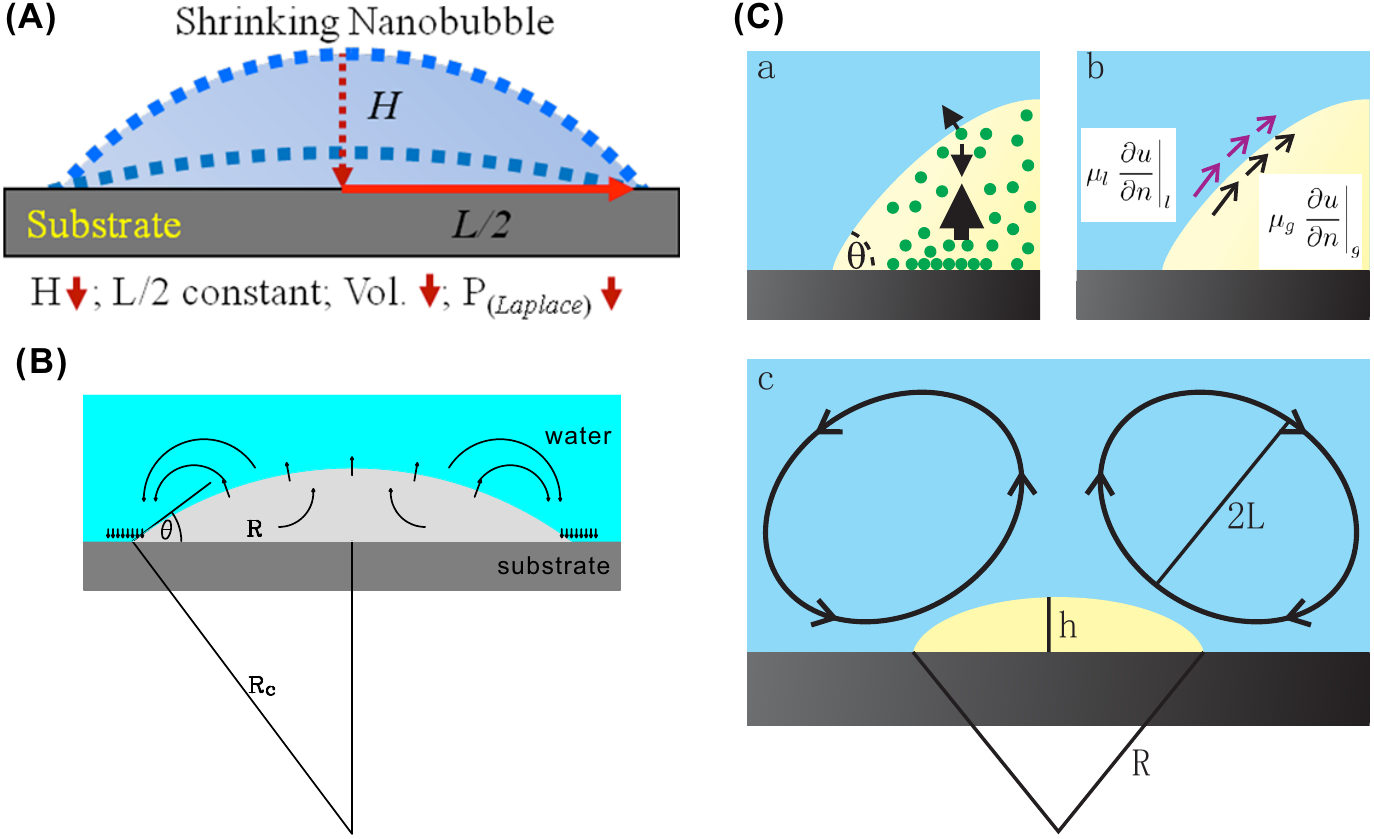}
    \caption{Schematic representation of main theories explaining the stability of nanobubbles. (A) Knudsen gas streaming. Upward flow of the Knudsen gas in the bulk, tangential component of the bulk Knudsen gas flow, and gas-rich liquid stream circulating around the bubble, which eventually transports the diffusive outfluxing gas back to the three-phase line for reentry. Reprinted with permission from Ref.~\cite{Seddon2011}; (B) Dynamic equilibrium mechanism for surface NB stabilisation. Sketch of gas influx and outflux that leads to the stability of NB. Reprinted with permission from Ref.~\cite{Brenner2008}; (C) Contact line pinning during NB growth and shrinkage. Reprinted with permission from Ref.~\cite{Zhang2013}. Copyright 2013 American Chemical Society.}
    \label{fig:fig-theory}
\end{figure}

Dynamic equilibrium is another theory to explain the stability of surface NBs. It suggests that a continuous influx of gas near the contact line
be able to sustain and stabilise the NBs gas phase in contact with
hydrophobic substrates (see Fig.~\ref{fig:fig-theory}B) \cite{Brenner2008}.
According to this theoretical model, the substrate hydrophobicity leads to the
attraction between the gas and the substrate, which results in a gas influx that balances an expected
outflux due to the Laplace pressure,
in this way establishing a dynamic equilibrium \cite{Dammer2006,Zhang2007}.
The model also predicts the equilibrium radius of the NBs and a threshold for gas concentration
and substrate hydrophobicity that can enable such dynamic equilibrium \cite{Brenner2008}.
In particular, the stability of NBs is observed for a narrow temperature and dissolved-gas concentration
range \cite{Seddon2011a}. Moreover, it has been suggested that NBs are of
Knudsen type in order to be stable in view of the thermal
energy of the substrate and the diffusive nature of the liquid--gas interface (see Fig.~\ref{fig:fig-theory}C \cite{Seddon2011}.
Due to the gas flow, a related liquid flow as a circulatory stream
from the apex to the three-phase contact line would manifest itself and
would further guarantee the balance between gas influx and outflux, in agreement with
experimental observations \cite{Seddon2011a}.
The dynamic equilibrium model
has predicted an upper bound for the NB size, which is also in agreement with experiments \cite{Seddon2011a}.
To this end, the maximum and minimum bubble size decreases with temperature \cite{Petsev2013},
while the gas-side contact angle has been found to decrease with decreasing NB size \cite{Kameda2008,Kameda2008a}.

Finally, classical Density Functional Theory (DFT) as a numerical and theoretical method has contributed to a better understanding of the
thermodynamic properties of surface NBs.
In particular, constrained Lattice Density Functional Theory (LDFT) and kinetic LDFT
suggest that NBs exist in a thermodynamic metastable state rather than at a thermodynamic
equilibrium \cite{Liu2013b}, in agreement with recent simulation studies \cite{Che2017}.
DFT could readily provide relationships for the gas-side contact angle,
size, and chemical potential of NBs,
suggesting that NBs stability stems from contact line pinning, which
is a result of oversaturation
and substrate roughness due to chemical heterogeneities.
Also, substrate hydrophobicity has been found to favour the stability of NBs \cite{Liu2013b}.
These findings are in agreement with a series of recent experimental findings in the area of NBs, which
will be discussed in more detail in subsequent sections.

%Mean field results.\cite{Bauer1999,Snoeijer2008}    ----> DFT and asymptotic analysis of the contact angle (Generally about contact angles and NOT nanobubbles)

\subsection{Simulation}\label{sec:sec2-2-simulation}

\begin{figure}
    \centering
    \includegraphics[width=\columnwidth]{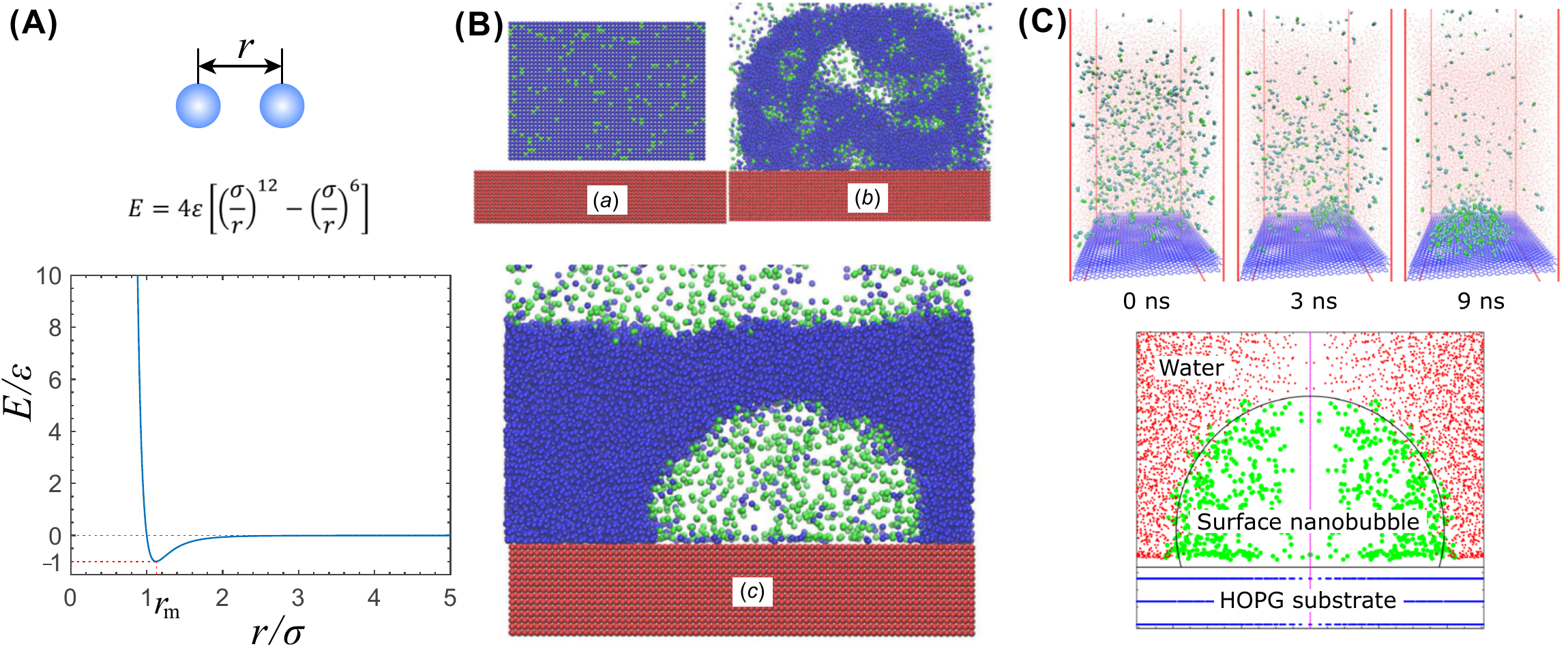}
    \caption{MD simulation based on LJ potentials (all-atom and CG models).
    (A) LJ potential as a function of distance $r$ between CG beads or atoms;
    (B) NB formation from MD simulation of a CG model. Reprinted with permission from Ref.~\cite{Weijs2012c};
    (C) NB formation from N$_2$ and O$_2$ atoms on an HOPG substrate immersed in water from all-atom MD simulation. Reprinted with permission from Ref.~\cite{Che2017}.
    }
    \label{fig:fig03-md}
\end{figure}

Molecular dynamics (MD) simulation has been used to study surface NBs,
due to its flexibility in providing a molecular-level description
in dense systems and the availability of reliable force-fields and highly parallelised free simulation software for both all-atom and
coarse-grained (CG) models (see Fig.~\ref{fig:fig03-md}).

By means of MD simulation,
Weijs \textit{et al.}\ investigated the formation and stability mechanisms of bulk NBs
in a binary mixture of simple Lennard--Jones (LJ) CG particles (see Fig.~\ref{fig:fig03-md}B) \cite{Weijs2012d,Weijs2012c}.
The MD simulations were able to capture the formation and growth of NBs and
estimate their final size allowing for a direct comparison of the results with
theoretical predictions from nucleation theory. The authors concluded that
NBs are stable as far as they are at a large enough
distance between them so as to protect themselves from diffusion by a
shielding effect, which is in agreement with simple diffusion calculations
and experimental results based on noncoalescence observations for macroscopic
bubbles \cite{Weijs2012d}. Further work by Weijs \textit{et al.}\ with  MD simulation has underlined the
importance of the hydrophobic nature of the
substrate in order to favour the spontaneous formation of NBs \cite{Weijs2012a}.
In this case, the observations from CG MD simulation align with the theoretical expectation
of the dynamic equilibrium model, which suggests that NBs be stabilized by a
nonequilibrium mechanism \cite{Weijs2012c} requiring a balance of gas influx and outflux \cite{Brenner2008}.
This CG model has
provided LJ parameters that can capture the formation of NBs and can be used for
further investigations based on NBs modelling \cite{Weijs2012a}. For example,
Maheswari \textit{et al.}\ have studied the stability and growth of NBs on chemically
patterned substrates by using MD simulations based on the latter model \cite{Maheshwari2016}.
The simulations revealed the importance of contact-line pinning and gas oversaturation
for NBs stability.
The latter studies have been recently extended to consider the case of two neighbouring
surface NBs on a chemically heterogeneous substrate \cite{Maheshwari2018}. In this case,
a diffusion equation-based stability analysis suggested that the NBs remain stable
when the contact line is pinned with their radii of curvature being equal \cite{Maheshwari2018}.
Li \textit{et al.}\ have systematically studied the coalescence of
two neighbouring NBs by means of CG MD simulation and analysed their results
on the basis of different ratio of interactions between gas,
liquid and substrate \cite{Li2018}. When the contact line is not pinned, different
coalescence scenarios are possible, such as direct merging through the existence
of a film layer of gas molecules between the NBs.
By means of MD simulations, Hong \textit{et al.}\ \cite{Hong2019} predicted
that NBs of a certain radius $R$ are stable
when the distance between bubbles is smaller than a length scale $L$ proportional to
$R^{4/3}$, in this way providing an indication for the stable bubble concentration
in solution. 

MD simulation has also demonstrated that surface NBs can be stabilised
in superheated or gas oversaturated liquid due to the contact-line pinning that stems
from surface heterogeneity \cite{Liu2014b}. Both the pinning and the oversaturation or superheating
of the gas phase in liquid--vapour mixtures is a requirement for the stability
of surface NBs according to this MD study, which is also currently a dominant
assumption in the literature. Moreover, stable NBs are obtained at
moderate oversaturation when the radius of curvature and the contact angle of
the NBs decrease, whereas a liquid--vapour transition takes place at high
oversaturation (supersaturation) \cite{Liu2014b}. The simulations also predicted
that both the gas-side radius of the curvature and the gas-side contact angle increase as
the superheating/supersaturation levels increase.
However, NBs become unstable at high levels of superheating/supersaturation \cite{Liu2014b}.

The above assumptions have been supported by
recent all-atom MD simulation, which has also provided details on the formation, dissolution
and properties of surface NBs on an ideal HOPG substrate (see Fig.~\ref{fig:fig03-md}C) \cite{Che2017}.
NBs favourably formed on the substrate underlining
the importance of the substrate and its hydrophobic nature that leads to the
attraction of the air molecules.
While NBs were formed through a nucleation process, they
eventually dissolved, in this way providing evidence that
their stability can be attributed to other mechanisms (\textit{e.g.}\ contact-line pinning).
Most importantly, this study has provided first-hand measurements for NBs density
concluding that NBs consist of dense gaseous phases, which has been recently validated
by experimental data \cite{Wang2019}. Moreover, the gas-side
contact angle of NBs (see Fig.~\ref{fig:fig1}) is smaller than that in nanodroplets (liquid-side) \cite{Che2017}.
This difference in contact angle between bubbles and
droplets with size smaller than about 10~{\textmu}m
has been recently studied by Zhang and Zhang
by means of theory and MD simulation of a CG model \cite{Zhang2019}. They
found that this mismatch becomes more pronounced as the size of these nano-objects
decreases. Moreover, the contact angle in the case of nanodroplets is
size and model dependent \cite{Theodorakis2015}.
In addition to the contact-angle mismatch between nanodroplets and NBs,
the all-atom MD simulation has found that the surface tension in NBs is smaller
than the corresponding surface tension of
a water--air interface under atmospheric conditions \cite{Che2017}. A similar model has recently focused
on the mechanical stability of surface NBs by Dockar \textit{et al.}\ \cite{Dockar2018}. In this study,
quasi-two-dimensional and three-dimensional nitrogen surface NBs were investigated and new
cavitation threshold models were proposed and assessed. This study suggests that the
discrepancies between experiments and the Blake threshold are due to the contact-line pinning,
in this way providing further evidence for pinning as a plausible mechanism for stabilising surface NBs \cite{Dockar2018}.
By means of all-atom MD simulation Chen \textit{et al.}\ have recently
investigated the stability of surface NBs on hydrophobic surfaces \cite{Chen2018}.
They found that NBs were stable for longer than 160~ns and
the stability depends on the gas adsorption, the solid--gas interaction
energy, and the bulk gas concentration. Moreover, an increased hydrophobicity of the
substrate alleviates the requirement of oversaturation conditions. The authors also
suggest that the gas enrichment layer, the gas adsorption monolayer on the
substrate, and the water hydrogen bonding near the interface could be the necessary conditions
for the stability of NBs \cite{Chen2018}. In contrast, they concluded that three-phase pinning sites are not necessarily required.
All-atom simulation has also explored the
midrange nanoscale hydrophobic interaction by NB nucleation
between nanometre-sized hydrophobes \cite{Koishi2004}. Koishi
\textit{et al.}\ \cite{Koishi2004} found that the NB formation exhibits hysteresis
when the size of the hydrophobe is larger than 2~nm. By means
of potential-of-mean-force calculation, they were able to provide
an estimation of the strength of the nanoscale hydrophobic
interaction.

Given that the solvent exchange procedure (discussed in Section \ref{sec:sec4-formation}) has become a widely used protocol
to produce surface NBs, MD simulation has attempted to study this process as well \cite{Xiao2017}.
In a recent study, Xiao \textit{et al.}\ found that a solvent--solvent interface
that traps gas molecules forms during the solvent exchange process \cite{Xiao2017}.
As this interface moves against the gas
concentration gradient and towards the substrate, the local oversaturation
of gas molecules initiate the nucleation of the NB on the substrate or in the bulk solution.
Xiao \textit{et al.}\ were able to draw a phase diagram of substrate wettability \textit{versus}
gas oversaturation, which indicates areas of favourable NBs formation \cite{Xiao2017}.
Moreover, Xiao \textit{et al.}\ have carried out further MD investigations to
assess the effect of surfactants on the stability of
surface NBs \cite{Xiao2017a}. Their simulations suggest that the presence of surfactants may lead
to the loss of stability of the NB as a result of either the
adsorption of surfactant onto the substrate or the decrease of the vapour--liquid
surface tension. This would lead to the depinning of the contact line,
especially when the surfactants are water-insoluble.
It is anticipated that this study may have further implications regarding the effect of other contaminants on the
stability of gaseous phases at liquid--vapour interfaces.

A very recent work by Molinero \textit{et al.},
which highlights the predictive capabilities of CG MD simulation, has
focused on the nucleation mechanisms as well as the formed stationary states of
electrochemically generated NBs \cite{Sirkin2019}. The MD simulation is able to
capture the molecular pathway of NB nucleation toward formation on nanoelectrodes and
characterise the stationary states. In particular, the simulation indicates a classical
mechanism for NB nucleation, where different nucleation regimes depend on the binding
free energy per area of the NB to the electrode. This work predicts different states for the
NB, such as micropancakes attached to the electrodes and homogeneously nucleated NBs close to
the electrode without attachment. The authors conclude that NBs nucleate
heterogeneously. Moreover, the stronger the driving force for the electrochemical reaction,
the larger the size of the NB and the higher its contact angle with the electrode.

Finally, Many-body Dissipative Particle Dynamics (MDPD) simulation has explored
the sliding dynamics of a NB on a surface indicating that the surface roughness and
the bubble shape affect NB motion  \cite{Wu2017}. For small gas-side
contact angles, NB mobility is higher even than the rising of
a bulk NB, while below a contact-angle threshold-value, the mobility of a NB
on a substrate with roughness can be higher than the mobility of a NB on a flat (smooth)
substrate due to the superlyophobicity, which can reduce the friction resistance \cite{Wu2017}.

\subsection{Experiment}\label{sec:sec2-3-experiment}
%Most of the work on surface NBs relates to experimental studies.
Before considering surface NBs, it is
worthwhile to briefly discuss the stability of free (bulk) NBs \cite{Alheshibri2016}.
In fact, much literature refers to bulk NBs, which
can form spontaneously under different salt concentration and pH \cite{Jin2007}.
Bulk NBs have been generally found to be stable in the case of alkaline solutions,
in contrast to what has been observed in solutions with high ionic strength \cite{Jin2007}.
Moreover, the zeta-potential has indicated that NBs have an electric double layer of negative charge,
which is attributed to the adsorption of OH$^-$ at the gas--water interface.
This electric double layer may prevent the aggregation of bubbles and reduce the surface
tension, which, in turn, may decrease the internal pressure in the bubble. Moreover, the
charged interface may lead to additional Maxwell pressure and further decrease of
the surface tension, which may explain the reason that NBs are
stable in aqueous solutions of water-soluble organic molecules of low molar mass \cite{Jin2007}.
Bulk NBs have been also studied in ultra-pure water solutions of LiCl and NaCl by means of
Rayleigh/Brillouin scattering, highlighting a dependence
of NB stability on salt concentration \cite{Duval2012}.
Long lifetime and ageing effects in NBs
were attributed to the electric charge, while their stability might be due to the
coverage of their surface with negative ions \cite{Duval2012}.

\begin{figure}
    \centering
    \includegraphics[width=0.65\columnwidth]{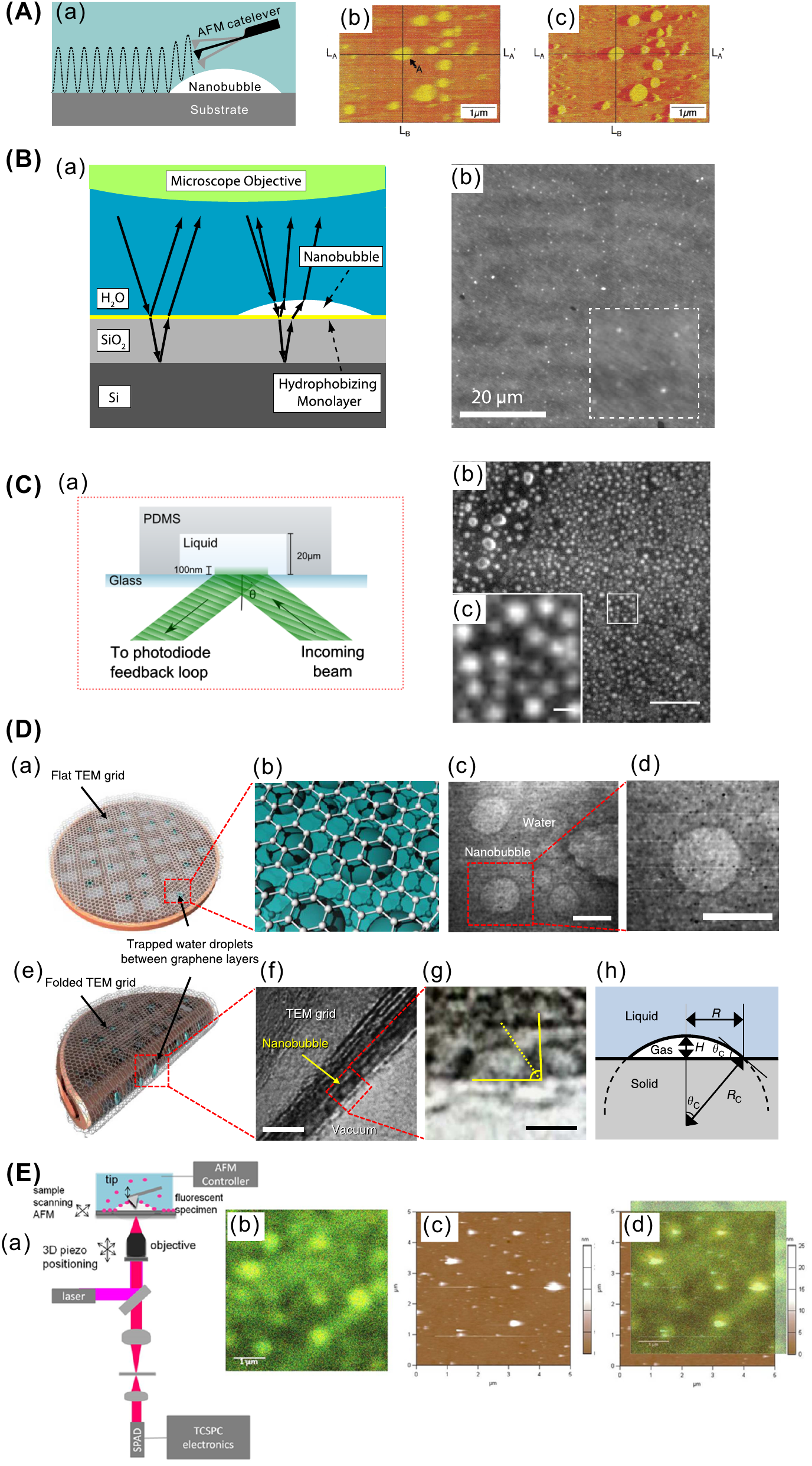}
    \caption{Main experimental methods for NBs imaging:
    (A) Schematic illustration of TM-AFM and first AFM obtained images. Panels A(b-c) are reprinted with permission from
    Ref.~\cite{Ishida2000}. Copyright 2000 American Chemical Society;
    (B) Schematic diagram of the nonintrusive optical interference-enhanced reflection microscopy
    technique and example of optical imaging by using this method. Reprinted with permission from
    Ref.~\cite{Karpitschka2012};
    (C) Total-internal-reflection-fluorescence (TIRF) microscopy for the study of NB dynamics.
    Schematic illustration of the TIRF microscopy for NB measurement
    %Cross-section of the polydimethylsiloxane channel for studying the water--ethanol--water exchange
    and nanobubbles observed under TIRF microscopy. Reprinted with permission from Ref.~\cite{Chan2012};
    (D) Ultra-high vacuum TEM. Graphene liquid cell for TEM measurement and TEM images showing the morphologies of NBs in the
    graphene liquid cell. Reprinted with permission from Ref.~\cite{Shin2015};
    (E) Schematic representation of fluorescence lifetime imaging microscopy and obtained results
    of surface nanobubbles nucleated by an ethanol--water exchange on OTS/glass. Reprinted
    with permission from Ref.~\cite{Hain2016}. Copyright 2016 American Chemical Society.
    }
    \label{fig:fig-experiment}
\end{figure}

In the case of surface NBs, Ishida \textit{et al.}\ have carried out experiments on surface NBs by using tapping-mode
AFM (TM-AFM) \cite{Ishida2000} (see Fig.~\ref{fig:fig-experiment}A). In this case, a single wafer, hydrophobised with
OTS, was immersed in water.
The sensing mechanism of NB in TM-AFM is based on the repulsive force that produces
a bump in the image, assuming that the AFM tip does not penetrate the bubble  \cite{Lou2000}.
The height of the NBs was determined on the phase image and the force curves and
the apparent contact angle (gas side) was estimated to be much smaller than the
macroscopically expected contact angle. At that time, the small gas-side contact angle
was
believed to be one of the reasons for the stability of surface NBs \cite{Ishida2000}, while theory currently suggests that
surface NBs with gas-side contact angles smaller than 90$^\circ$ can
be stable \cite{Chan2015a}.
Further studies by TM-AFM have revealed close-packed NBs with irregular cross-sections
and radii of curvature about 100~nm and height 20--30~nm \cite{Tyrrell2001,Attard2002}.
NBs on a hydrophobic glass for a range of different
pH conditions indicated that the increase of pH leads to smaller and more
uniform NBs, which was attributed to the presence of the surface charge
as in the case of bulk NBs \cite{Tyrrell2002}.
TM-AFM studies on atomically flat substrates have observed stable surface NBs for
hours during experiments with the smallest NBs being around 10~nm.
Further experiments of surface NBs, which were formed for the first time
by using the ethanol--water exchange method (nowadays a standard method to generate surface NBs discussed later in Section \ref{sec:sec4-formation})
on an OTS silicon hydrophobic substrate, have
found results that are consistent with experiments of NBs on HOPG substrate \cite{Zhang2006}.
By considering a wide range of solutes, such as  multivalent salts, cationic,
anionic, and nonionic surfactants, and solution pH, Zhang \textit{et al.}\ concluded that
these factors influence very little the morphology of NBs \cite{Zhang2006}.
In agreement with previous results, the gas-side contact angle of NBs was estimated to be smaller than
the expected macroscopic angles on the same substrate suggesting that the Laplace
pressure may be smaller, since a larger radius of curvature would imply a
small pressure \cite{Zhang2006}. Still, the predicted theoretical value of the pressure
is smaller than the one obtained from experimental measurements \cite{Zhang2006}.
In the so-called force modulation mode of TM-AFM, which is based on
the interaction between the cantilever tip and the bubble,
the height and the adhesive force for the NB has been obtained \cite{Bhushan2008}.
By using a viscoelastic model, the mechanical properties of NBs can be assessed by
means of the stiffness and the damping coefficient.
The modes were also set to study the effect of the surface tension
on the attractive interaction force and the contact angle hysteresis during
the tip--bubble interaction \cite{Bhushan2008}. Finally, Wang \textit{et al.}\ \cite{Wang2009} found that nanoindents are formed after immersing an ultrathin polystyrene film in water,  and the size and location of the nanoindents are strongly correlated with
that of NBs.

TM-AFM experiments have also been used to study the formation propensity of NBs and their
distribution on an ultra-thin polystyrene (PS) film in deionized (DI) water and
saline (sodium chloride) solution as a function of electrolyte, roughness, pH, and
substrate bias \cite{Mazumder2011}. In this study, the saline environment favoured
NBs of larger size in comparison with NBs in DI water. Increased roughness also led to
larger bubbles. The study also confirmed previous results (for both bulk and surface NBs)
that NBs are more
stable in alkaline solutions than in acidic. Finally, the size of NBs was found to
increase with positive bias \cite{Mazumder2011}. TM-AFM experiments have also considered
the stability and coalescence of NBs with lateral size from 100~nm up to around 10 {\textmu}m
and height from 10~nm to 300~nm on PS--water interface \cite{Li2014}. It was found that
the number of gas molecules increased by 112.5\% after coalescence, which was attributed
to the gas influx coming from the depinning of the contact line and the decrease of the
inner pressure during coalescence \cite{Li2014}. To this end, the coalescence was slower
for larger bubbles. The results are consistent with the contact line pinning theory and
experimental studies, which have estimated the lifetime to be about 6.9 hours in this case \cite{Li2014}.
Hence, this study has provided support in favour of contact-line pinning, gas influx near
the contact line and a thin `contaminant film' around the gas--liquid interface,
and even the electrostatic effect, in this way incorporating many elements of
previous theories.

NBs on hydrophilic substrates immersed in water, such as self-assembled monolayers (SAMs) \cite{Song2016a},
have been observed by \textit{in situ} AFM for the first time by Song \textit{et al.}\
with gas-side contact angles larger than 94$^\circ$ \cite{Song2011}. In this case, the size
of the NBs was found to depend on the composition of SAMs. By fitting the
NB shape to a spherical cap, the height and the radii of the basis and the curvature
were found in agreement with previous studies. In the case
of hydrophilic SAMs, the macroscopic and microscopic contact angles also show agreement \cite{Song2011}.
On the contrary, in the case of smooth hydrophilic dehydroxlylated silicon
oxide wafer surfaces in water, TM-AFM experiments suggest that the formation
of NBs is not possible \cite{Yang2003}. However, randomly distributed NBs
were observed on methylated surfaces with controlled roughness, which, according to the authors,
may indicate that the pinning of the contact line
can stabilise the NB allowing for a small microscopic gas-side contact angle \cite{Yang2003}.

It should be noted that the invasive nature of AFM certainly affects the apparent shape of NBs.
For this reason, Walczyk \textit{et al.}\ have studied NBs on HOPG substrates in water by using
both TM-AFM and PeakForce (PF) mode AFM \cite{Walczyk2013}. In the case of PF-AFM, the force
exerted on NBs was measured instead of the resonating cantilever with the apparent size and
shape depending on the force. Even for forces as small as 73~pN, the true size of the NBs appeared
to be smaller in the experiment with the error in the estimation
increasing for larger NBs \cite{Walczyk2013}.
The height images obtained by PF-AFM indicated a decreasing apparent size of NBs with
increasing scanning force, which has been also observed in TM-AFM experiments, but there are
differences in the absolute values of the contact angles \cite{Walczyk2013}. Results of
PeakForce quantitative Nano-Mechanics (PF-QHM) measurements in the case of HOPG substrates,
which were also confirmed with TM-AFM experiments, estimated the stiffness of the NB from
60 to 120~pN/nm, with smaller NBs being stiffer \cite{Zhao2013}. Moreover, results
on the morphology between the PF-QHM method and TM-AFM were consistent \cite{Zhao2013}.
The contact angle of NBs on HOPG in water has been estimated about $61\pm4^\circ$ (gas side) by TM-AFM
provided that the cantilever is clean and the HOPG
substrate smooth, otherwise angles (gas side) as low as 30$^\circ$ can be observed due to the contamination that presumably introduced roughness on the HOPG substrate \cite{Borkent2010, An2015}. Such contamination effects also affect
the formation of NBs having provided further evidence for the contact-line pinning
hypothesis for the stability of surface NBs \cite{Berkelaar2014}.
In general, contamination issues are critical in
the interpretation of NBs experiment.
Berkelaar \textit{et al.}\ have convincingly pointed out that
NB-like objects can be induced by the use of disposable needles
in which PDMS contaminated the water. Therefore, nano-objects that
look and behave as NBs, in some cases they might simply be induced
by contamination \cite{Berkelaar2014}. The effects
on the contact angle of NBs stemming from different contamination
scenarios have been discussed in detail by
An \textit{et al.}\ \cite{An2015}.

Still, the above results should be put in perspective considering the effects of
different cantilevers and AFM modes on the experimental results.
In particular, it is widely accepted that experimental measurements of the shape and size of NBs depend on the cantilever properties. This has been
concluded by a study where 15 different cantilevers were used \cite{Borkent2010}.
Moreover, three different AFM modes, namely, tapping-mode, lift-mode, and force-volume-mode (FVM)
were tested in the case of HOPG substrates in water \cite{Walczyk2014}.
It was found that the tip--bubble interaction
strength depends on the position of the tip relative to the NB.
The direction of the tip movement was also found to have hydrodynamic effects, and hence influences the deformation of the NB \cite{Walczyk2014}.
The hydrodynamic effects were more pronounced in the FVM, where the tip approaches the NB from above,
In contrast, the hydrodynamic effects were less pronounced  in the lift-mode, where the tip approaches the NB from the side.
In addition to the tip shape and cleanness, the scanning model in AFM
experiments is another factor that influences the morphology observations, especially when the tip
movement is complex \cite{Walczyk2014}. By combining TM- and FVM-AFM methods with hydrophilic
and hydrophobic AFM tips, Walczyk \textit{et al.}\ analysed the NB deformation and
its interaction with respect to the tip position on the NB \cite{Walczyk2014}. The results
indicated that the tip--bubble interaction strength and the magnitude of the bubble
deformation depend on the vertical and horizontal positions of the tip on the bubble
in the case of hydrophobic tips \cite{Walczyk2014}. Moreover, hydrophobic tips
led to severe bubble deformation due to the permanent contact of the tip
with the NB, which has also led to the stretch of the NB
as the tip was moving away from the bubble centre.
In contrast, a hydrophilic tip with no direct contact between the tip and the NB may reduce these
effects, in this way allowing for reliable AFM images and NB dimensions measurements \cite{Walczyk2014}.
In this case, a thin liquid film formed between the tip and the bubble. The deformation of the
NB in the vertical direction depended on both the vertical and the horizontal position of the tip,
independently of the tip hydrophobicity.
The deformation increased with decreased tip--sample separation
distance, with a faster decrease in the periphery of the bubble.
A flat profile for
the NB was obtained, which suggested a Laplace pressure closer to values of the
atmospheric pressure. Walczyk \textit{et al.}\ suggested that these effects and
the contact line pinning may explain the long lifetimes of NBs \cite{Walczyk2014a}.
In a recent study by Wang \textit{et al.},
the contact angle measurement of NBs on mica and molybdenite
substrates was investigated and its dependence
on the curvature radius of the AFM probe tip was discussed \cite{Wang2019b}.
The authors concluded that the true
contact angle is not always obtuse and largely affected by the tip curvature it
is rather closer to the microscopic value. 

Despite all the efforts by means of AFM experiments that have provided important
information on the morphology of NBs due to the very good spatial resolution, AFM
has certain drawbacks, such as the intrusiveness and the inability to provide
information on the chemical identity of the NBs as AFM is not able
to distinguish between NBs and other objects. Another disadvantage is the time-consuming
full-size images and the inability to capture any transient effects, such as the formation
of NBs, and the requirement of a mechanically and chemically stable environment
for the measurements. These are a few of the reasons that additional methods
have been employed individually or in combination with AFM techniques
to complement the study of surface NBs (see Fig.~\ref{fig:fig-experiment}).

In one of these methods, Attenuated Total Reflection Fourier Transform Infrared (ATR-FTIR)
and AFM can measure the gas transfer by using the infrared active CO$_2$ gas \cite{German2014}. The measurements show that
the NB gas exchanges with the dissolved gas in the liquid phase.
%and that the gas transfer is measurable by using the infrared active CO$_2$ gas \cite{German2014}.
While CO$_2$ NBs eventually dissolve, they remain stable for hours and the dissolution
rate depends on the initial size of the NBs. In this study, pinning of the contact line
was underlined as the main factor for the stability of surface NBs \cite{German2014}.
In another study, the pinning of NBs was again highlighted as the main
reason for NBs stability by using AFM, optical microscopy and a high-speed camera
in the case of OTS glass at 37$^\circ \rm C$ \cite{Zhang2014}.

Nonintrusive interference-enhanced reflection microscopy has also been used to visualise
individual NBs (see Fig.\ \ref{fig:fig-experiment}B), which has shown that their formation is not provoked by the
intrusive nature of the AFM technique \cite{Karpitschka2012}, contrary to
some experimental assumptions \cite{Walczyk2013}. Moreover, the growth
dynamics of NBs can be investigated observing stable NBs with gas-side contact angles up to 35$^\circ$.
This method combined with particle tracking techniques can describe the flow of liquid in the
vicinity of the NBs \cite{Karpitschka2012}.
Furthermore, the total-internal-reflection-fluorescence excitation method (see Fig.\ \ref{fig:fig-experiment}C) has been used to track the
NB formation on a hydrophilic glass during water--ethanol exchange resulting in
strongly contrasting images with high spatial resolution \cite{Chan2012}.
The nucleation dynamics during
the solvent exchange procedure was resolved and a Brownian motion was observed for
tracer particles near the NBs \cite{Chan2012}.

\textit{In situ} Transmission Electron Microscopy (TEM) has been used to investigate electrolytically
generated H$_2$ NBs, where the gas initially dissolved in the solution and then nucleated near Au
electrodes \cite{Liu2014}. The growth dynamics of the NBs indicated a dependence on the substrate roughness, while
dewetting (dissolution) appeared to be driven by a wetting instability \cite{Liu2014}. The growth
dynamics of small bubbles were not affected by neighbouring NBs, contrary to what
may happen for larger NBs \cite{Huang2013}. In the case of graphene liquid cells
(water encapsulated by graphene membrane), NBs were investigated by \textit{in situ}
Ultra-High Vacuum Transmission Electron Microscopy (UHV-TEM, see Fig.\ \ref{fig:fig-experiment}D) indicating two distinct
growth mechanisms for NBs, which depend on their relative size and the existence of
a critical radius \cite{Shin2015}.
In fact, the liquid cell electron microscopy is a new technique for \textit{in situ} imaging and control
of nanoscale phenomena to study nanoscale processes in liquids \cite{Grogan2014}.
Results have shown
that radiolysis is generally important and hydrogen and hydrated electrons can achieve
equilibrium concentrations within seconds,
while heating is typically insignificant \cite{Grogan2014}. A significant advantage of this method
is the ability to image bubble nucleation, growth and migration.
Based on a simplified reaction--diffusion model,
Grogan \textit{et al.}\ could predict the conditions for the formation of H$_2$ bubbles \cite{Grogan2014}.

The chemical identity, phase state, and density of NBs can be analysed
by means of Infrared Spectroscopy.
This method has been applied in the case of CO$_2$ NBs \cite{Zhang2007}.
The gas phase for different gas solubilities is in
atmospheric pressure, which might explain the high stability of NBs \cite{Zhang2007}.
Results coming from Attenuated Total Reflection Infrared Spectroscopy in
the case of CO$_2$ estimated the dimensions of bubbles in the
range 5--80~nm,  with bubbles being stable for 1--2 hours \cite{Zhang2008b}.
In this case, the pressure of the gas phase was estimated to be similar to
the ambient pressure. Further AFM experiments, which calculated the pressure
on the basis of the radius of curvature and the bubble dimensions agree with
the results based on the IR spectrum. Moreover, the lower pressure of CO$_2$ in
atmosphere and its greater solubility in water in comparison with N$_2$ and O$_2$
have been indicated as the main factors for explaining the lower stability in the case of CO$_2$ NBs \cite{Zhang2008b}.
Smaller NBs were found to have shorter lifetimes, while average pressure and
curvature of NB decreased with time. Finally, experiments of plasmon resonance provide
evidence that NBs are in gas phase and their low gas-side contact angle suggested that an
attractive force acts between the solid--air and liquid--air interfaces \cite{Zhang2008b}.
Previous Surface Plasmon Resonance (SPR) experiments in the context of a film
had confirmed a low refractive index at the interface with the average pressure being estimated
about 1~atm, which is generally consistent with the observed radius of
curvature ($R\sim4${\textmu}m) in NBs and possibly explains their long lifetime \cite{Zhang2007}.

Quartz Crystal Microbalance (QCM) can be used to investigate the kinetics of adsorption of CO$_2$ molecules dissolved in water on a
hydrophobised silica surface and results have been compared with those obtained
by AFM experiments \cite{Yang2003,Yang2007}. The results indicated an early adsorption of the gas ($<$20 min)
before this could be detected by TM-AFM. Hence, the sensitivity to detect low surface
coverage of NBs was much greater in the case of QCM than in the case of AFM,
in this way allowing for a more detailed study of the kinetic process of NBs formation \cite{Zhang2008}.
The process consisted of two different consecutive stages, namely a slow and a fast
adsorption process. The slow process was associated with the diffusion of gas molecules
from the interfacial region to the surface (Harvei nuclei). After a NB reached a critical
size, the gas adsorption took place through diffusion from the interfacial regions towards the
NB. Results from QCM experiments on bare and thiol-coated gold surface by using the
solvent-exchange method showed that the formation of NBs takes place
within less than a minute \cite{Zhang2008}. In this context, QCM has also highlighted possible applications of NBs
in efficient cleaning of solid--liquid interfaces to remove
bovine albumin, which takes place in a 10-second treatment, %and may be an
%environmentally friendly solution,
in contrast to the treatment with
SDS (sodium dodecyl sulfate) surfactants that requires about 20~min \cite{Liu2008}.

Finally, Small-Angle X-ray Scattering (SAXS) experiments have investigated the
formation of NBs on hydrophobic SAMs surface in a binary ethanol/water titration \cite{Palmer2011}.
SAXS revealed an electron density depletion layer at the hydrophobic interface with
changing air solubility in the immersing liquid due to the NB formation \cite{Palmer2011}.
Hence, NB formation was responsible for the so-called long-range hydrophobic force \cite{Palmer2011}.
Other methods to study NBs include the
Fluorescence Lifetime Imaging Microscopy (FLIM, see Fig.\ \ref{fig:fig-experiment}E) combined
with AFM to investigate the impact of surface treatment and modification on surface
NB nucleation \cite{Hain2016}. Also, Neutron Reflectivity measurements have been
used to measure the water density in the interface region \cite{Schwendel2003,Steitz2003}.

In summary, two different types of methods have been discussed here regarding the
experimental characterisation of surface NBs. The first type of methods is based on AFM, which
offers very high spatial resolution but low temporal resolution. Hence, dynamic
properties and the temporal evolution of the NB formation are challenging.
Moreover, AFM experiments do not distinguish between NBs and other nano-objects and
are generally unable to provide information on the chemical identity of NBs.
These are invasive methods in nature, which affect the shape of NBs and
results depend on cantilevers
(\textit{e.g.}\ hydrophobic or hydrophilic) and
AFM modes (\textit{e.g.}\ tapping, Peak Force, \textit{etc.}).
In addition, %full-size images are time-consuming without the ability of capturing transient
%effects, such as NB formation and growth.
AFM experiments also require stable chemical and
mechanical environments. The second type of methods is mainly optical
and non-intrusive. These methods offer lower spatial resolution than AFM, but can achieve
high temporal resolution. They offer continuous monitoring of relevant processes,
such as bubble nucleation and growth with
the ability to describe the kinetics of such processes. Combined
with particle tracking methods can describe the flow of liquid in the vicinity of NBs.
Finally, these methods are able to provide analysis on the chemical identity, phase state
and density of NBs. Non-intrusive methods with high spatial and temporal resolution
may constitute an ideal tool for future research in the area of surface NBs.

\section{Morphological characteristics}\label{sec:sec3-morphology}
The morphology of NBs is usually characterised by their size dimensions that determine the contact angle. Information about these properties is mainly obtained by AFM experiments and applying
fitting procedures \cite{Borkent2010,Walczyk2013}. In general, a spherical-cap shape
is implied to describe surface NBs with a circular three-phase boundary, height
of a few tens of nanometres and lateral extension of up to several microns \cite{Zhang2010}.
As a result, the contact angle (measured from the gas side) is expected to be small \cite{Borkent2010,Walczyk2013}.
Although the three-phase contact line of NB is usually circular, irregular (noncircular) shapes have been
observed \cite{Zhang2006}. Below, we discuss %results from different experiments and theoretical
%methods on a variety of substrates, including, also,
results related to the morphological and mechanical properties of NBs.

In fact, results on the contact angle of NBs (Fig.~\ref{fig:fig1}) vary and different values have been reported in the literature.
While early experiments have found values of about 20$^\circ$ on HOPG substrates \cite{Yang2007,Zhang2010},
Borkent \textit{et al.}\ have measured contact angles of
about 60$^\circ$ by means of AFM experiments \cite{Borkent2010}, which are among the highest values reported in the literature.
However, the contact angle is also expected to depend on NB's size
when its radius is smaller than $R=20$~nm \cite{Borkent2010}, which is in agreement
with previous AFM experiments \cite{Zhang2006}. When the roughness of the substrate increases, the contact angle
can obtain values around 30$^\circ$. Still, contact angles (gas side) measured by PFT-AFM indicate smaller values
(\textit{i.e.} between 5$^\circ$ and 35$^\circ$), which has been comparable with a range
5$^\circ$--20$^\circ$ measured more recently by TM-AFM \cite{Walczyk2013}.
The difference between microscopic and macroscopic contact angles on the same substrate
has been encountered by both experiment \cite{Zhang2006} and simulation \cite{Che2017},
where differences have been partly attributed to the line tension \cite{Ishida2000,Yang2003}.
In the case of nitrogen NBs with average diameters 10--100~nm on Au(111) substrates,
it has been experimentally shown that the line tension can change the sign from negative
to positive as the NB size decreases \cite{Kameda2008}, which underlines the role of the
line tension in the case of small NBs. This effect had been already confirmed in the very
first experiments as well as in the case of more recent experimental studies
\cite{Lou2002,Lou2000,Tyrrell2001,Simonsen2004,Borkent2010,Song2011,Zhang2012,Walczyk2013,Zhao2013}.
This suggests a larger radius of curvature in the case of NBs and a commensurate decrease in the
Laplace pressure \cite{Zhang2006}, which eventually leads to the modification
of Young's equation \cite{Boruvka1977,deGennes1985,deGennes2002}
in order to account for the influence of the line tension \cite{Amirfazli2004,Pompe2000,Herminghaus1999}.
In fact, Yang \textit{et al.}\ proposed that the
line tension might be responsible for the difference between
nanoscopic and macroscopic contact angles \cite{Yang2003}, which
has also been discussed in the context of nanodroplets \cite{Zhang2018, Fernandez2017,Good1979,Weijs2011}. Still, the experimental verification of the line tension
contribution is a challenging task \cite{Zhang2006,Yang2003}. By collecting data from about 200 NBs
at room temperature,
Zhang \textit{et al.}\ suggested that surface NBs on HOPG substrates that are larger than 100~nm in height and
2~{\textmu}m in curvature radius are rather unstable  \cite{Zhang2010}. On mica substrates, the radius of the basis was less
than 300~nm for air NBs, while air and hydrogen NBs on HOPG were in the range 50--450~nm \cite{Zhang2010}.
On heated substrates, NBs on HOPG  have exhibited larger lateral sizes of
about 8~{\textmu}m with all NBs having a similar aspect ratio independently of the
substrate temperature, which corresponded to a gas-side contact angle of 3$^\circ$--24$^\circ$ \cite{Xu2014}.

Results on a crystalline (111) Si wafer coated with a thick PS film (hydrophobic substrate) have
indicated that the radius of curvature of the NBs were of the order of 300~nm in this case \cite{Simonsen2004}.
Moreover, a spherical-cap shape has been assumed for the NBs and the apparent pressure
has been estimated as higher than the atmospheric pressure (about 4.5~atm) \cite{Simonsen2004}.
In similar studies by Tyrell \textit{et al.}, curvature radii of the order of 100~nm
and height 20--30~nm for NBs have been reported \cite{Tyrrell2001}.

In the case of OTS-coated substrates, the estimated gas-side nanoscopic contact angle of NBs was
about 12$\pm$9$^\circ$ and on HOPG 16$\pm$6$^\circ$, while the corresponding contact angles
for nanodroplets were 72$\pm$5$^\circ$
for OTS and 108$\pm$11$^\circ$ for HOPG \cite{Zhang2006}. The formation of NBs was facilitated
when the solvents (ethanol) and water were at 45$^\circ C$ than at room temperature \cite{Zhang2006}.
Finally, the average height of the NBs was 26~nm and the lateral size about 591~nm \cite{Zhang2006}.

The mechanical properties of NBs on HOPG can be mapped with high resolution by means of PF-QHM
(PeakForce Quantitative Nano-Mechanics) AFM \cite{Zhao2013}. The stiffness of the NBs lies
within 60 and 120 pN/nm. This behaviour was size-dependent with bigger NBs being softer than
the smaller ones \cite{Zhao2013}. Measurements from TM-AFM suggest that the NBs be harder than
the corresponding nanodroplets \cite{Kameda2008a}.

The density of nanobubbles has been estimated by means of all-atom MD simulation. The obtained value
was 409 kg/m$^3$ \cite{Che2017}, which is of the order of the density of liquids, in
agreement with recent experiments \cite{Wang2019}.
Moreover, all-atom MD simulation has found that the
surface tension in the case of NBs is about 50 mN/m, a value smaller than the water--air interface
tension at atmospheric conditions \cite{Che2017}.

\section{Formation of nanobubbles}\label{sec:sec4-formation}
Nanobubbles can form spontaneously by simple immersion of a solid
substrate \cite{Yang2003,Simonsen2004, Borkent2007,Yang2007,Seddon2011a,Berkelaar2012,Agrawal2005}
%with heights
%in the range 5--100~nm being
%reported in the literature
%\cite{Agrawal2005}.
Although the oversaturation of
dissolved gas is not a requirement for nucleation and formation of NBs,
a certain gas concentration (100--110\% gas concentration) and temperature
of the liquid (between 25$^\circ$ and 45$^\circ$) seems to favour
the formation of NBs \cite{Seddon2011}, while the range of
temperature for NBs formation appears to only depend weakly on the type of gas \cite{vanLimbeek2011}.

\begin{figure}
    \centering
    \includegraphics[width=0.5\columnwidth]{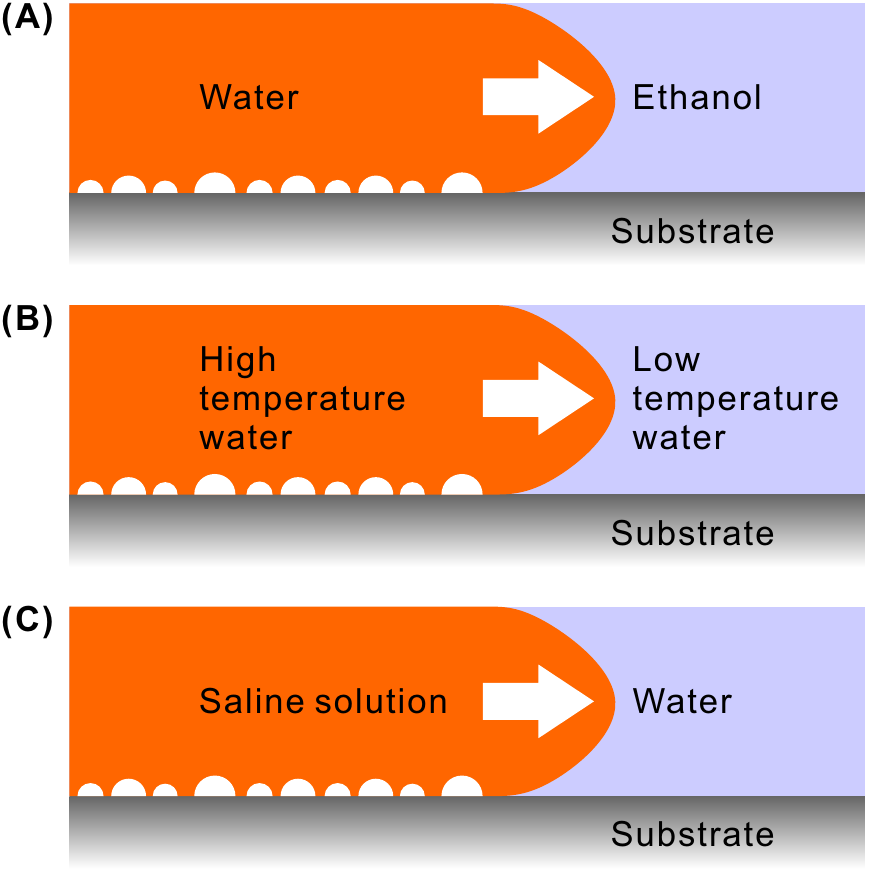}
    \caption{Formation of nanobubbles in experiment using the solvent exchange method. This can be achieved by using water--ethanol (A), temperature difference (B), and saline solution (C).}
    \label{fig:fig-formation}
\end{figure}

Despite the possibility of spontaneous formation,
the solvent exchange method is commonly used to form NBs (see Fig.~\ref{fig:fig-formation}A) \cite{Zhang2004,Zhang2005,Peng2016},
which has been applied in the case of substrates with different chemical and physical
properties
\cite{Martinez2007,Palmer2011,Yang2007,Yang2007a,Chan2012,Ishida2012,Karpitschka2012,Belova2013,Zhang2006,Zhang2011,Zhang2013a}.
In the case of growing microbubbles  on highly
ordered hydrophobic microcavity arrays with the solvent
exchange method, bubbles self-organise
into symmetric patterns, whereas asymmetric patterns were observed
in the case of larger distance between the microcavities \cite{Peng2016}.
NBs as small as 10~nm have been reported by using
the solvent-exchange method.
In this process, deionized water (DI water) is firstly injected into the liquid cell and then
replaced with pure ethanol \cite{Lou2000}. After waiting for several minutes, the ethanol
is replaced with DI water and many NBs can be produced.
%as has been shown, for example, in the case of mica substrates \cite{Lou2000}.
The underlying principle of this method is the gas separation during the mixing
process of the two liquids, due to its different solubility \cite{Lou2000}.
The advantage of this approach is that any other contamination is generally avoided.
To this end, 1-propanol has been mostly used in the solvent-exchange process by showing the
largest increase in range followed by ethanol and methanol \cite{Hampton2008}.
According to recent MD simulations, the oversaturation of the gas either
on the substrate or in the bulk solution, which depends on substrate hydrophobicity and the degree of
local gas oversaturation, may be the factor that favours the formation of the NBs
in the case of the solvent-exchange process \cite{Xiao2017,Che2017}.
Although the solvent-exchange method has high repeatability and
generally low or no contamination, AFM has imaged organic pollutants introduced by the alcohol, which
may even obscure the evidence of the existence of NBs \cite{Guan2012}. For this reason,
the Temperature Difference Method has been used to address such shortcomings
by avoiding the use of alcohol, as has been shown, for example, in the case of
NBs on HOPG substrates \cite{Guan2012}.
Here, low-temperature water is replaced by high-temperature water and NBs
form during the mixing process (see Fig.~\ref{fig:fig-formation}B) \cite{Guan2012,Guo2012}.
Finally, similarly to the solvent-exchange method, the saline solution/water exchange
method has produced NBs for different concentrations and valences of
saline liquids (see Fig.~\ref{fig:fig-formation}C) \cite{Liu2013a}.

Surface NBs can be also formed during photochemical or electrochemical processes.
For example, photocatalytic reactions can produce surface NBs by generating
hydrogen in methanol/water solution when
UV light illuminates a TiO$_2$ coated substrate \cite{Shen2008}.
The produced NBs existed during the photocatalytic reaction \cite{Shen2008}.
In the case of electrolysis of water to form NBs on HOPG substrates,
the HOPG substrate acts as a negative (positive) electrode to produce
hydrogen (oxygen) NBs with the coverage and size of NBs increasing with the voltage \cite{Yang2009a}.
In this approach, water with a small amount of sodium chloride allows for larger currents,
but results are similar as in the case of pure water \cite{Yang2009a}.
The existence of electrochemically generated NBs by
hydrogen gas on HOPG have
been initially confirmed by Zhang \textit{et al.}\
\cite{Zhang2006a}. It has been demonstrated that the formation and growth of NBs was
controlled by the applied voltage or the reaction time. Moreover, the authors
were able to observe the formation, growth, coalescence and eventual release of
merged NBs from the HOPG substrate \cite{Zhang2006a}.
Electrochemical generation of individual H$_2$, N$_2$, and O$_2$ NBs
have been reported as well in the literature \cite{Chen2015,Luo2013,Chen2014,Chen2015b,German2016, German2016b,Ren2017,Liu2017}.
In particular, Liu \textit{et al.}\ has demonstrated
by means of experimental and numeical methods
that a single NB at a Pt nanodisk electrode
is sustainable due to the H$_2$ electrogeneration
and better agreement between experiment and simulation is achieved
when a recessed electrode geometry is assumed \cite{Liu2017}.
In the case of individual O$_2$ NBs generated by
electrooxidation of hydrogen peroxide (H$_2$O$_2$),
a minimum concentration of  O$_2$ is required to observe
NB nucleation at the Pt electrode surface, which is about
130 times larger than the equilibrium saturation concentration of
O$_2$ \cite{Ren2017}.
Moreover, NBs can be generated for both
positive and negative potentials. In this work, Ren \textit{et al.}\
were also able to generate alternatingly single O$_2$ and H$_2$
NBs within the same experiment, in this way allowing for a direct
comparison of critical concentrations for nucleation \cite{Ren2017}.
German \textit{et al.}\ have provided measurements of the Laplace
pressure of single NBs with radius between 7 and 200~nm showing
a linear relationship between NB's Laplace pressure and its
reciprocal radius, in agreement with the classical thermodynamic
description \cite{German2016b}. German \textit{et al.}\ have also
measured the lifetime of individual
hydrogen and nitrogen NBs by using a fast-scan
electrochemical technique \cite{German2016}. In particular,
they found that the dissolution of NBs is partly limited by
the translocation of molecules across the gas--water interface,
where the interfacial gas transfer is estimated to be around
10$^{-9}$~mol/(N$\cdot$s) \cite{German2016}.
Chen \textit{et al.}\ have studied the nucleation and stability
of individual electrochemically generated H$_2$ and N$_2$ NBs
at platinum nanoelectrodes \cite{Chen2014,Chen2015b}.
In the case of H$_2$ NBs, similarly to O$_2$ NBs a large
concentration of gas at the electrode favours nucleation of NBs
\cite{Chen2014}. Moreover, the addition of surfactants would decrease
the nucleation barrier and an amount of surfactant will accumulate
at the H$_2$--solution interface hindering the transfer of H$_2$ molecules
to the solution. As a result, the residual current
reduces when the concentration of surfactant
increases \cite{Luo2013,Chen2014}. Moreover, Luo and White suggest
a two-step mechanism for the nucleation and growth of
NB \cite{Luo2013}.
In the case of nitrogen NBs \cite{Chen2015b}, the nucleation and
growth of a single NB was studied. It was also found that
the size of a stable NB nucleus depends on the concentration of
the gas at the electrode, which is two orders of magnitude
larger than the saturation concentration at room temperature and
atmospheric pressure \cite{Chen2015b}. Also, the residual current
after NB formation was proportional to the N$_2$H$_4$ concentration
and the nanoelectrode radius, which indicates that the dynamic
equilibrium depends on the N$_2$H$_4$ electrooxidation at the
three-phase contact line \cite{Chen2015b}. 
Spontaneous reaction between hydrogen and oxygen within bubbles of
diameter smaller than 150~nm has been also reported in the literature,
as a result of the high Laplace
pressure and fast dynamics of the gas atoms within the bubble \cite{Svetovoy2011,Svetovoy2014}.
In terms of applications, hydrogen NBs produced through
electrochemical reactions
can be further used as a template for the synthesis of metal hollow nanoparticles \cite{Huang2009} or nanoporous thin films \cite{Hui2009}.

\section{Stability: Contact-line pinning}\label{sec:sec-stability}
The stability of surface NBs is indicated by their lifetime.
A long lifetime may imply a duration from a few hours to many days, which is beyond the
expectation of the immediate dissolution of surface NBs due to the theoretically predicted
high Laplace pressure. To this end, the currently dominant explanation in the literature
for the high stability of surface NBs is the combined effect of
contact line pinning and gas oversaturation, which we will discuss here in more detail.

Initially, it was observed that the formation of NBs is sensitive to the surface topography,
as has been shown in the case of NBs on an HOPG substrate \cite{Yang2009}.
NBs were only formed at the upper side of the atomic
stops (hydrophobic areas), while no NBs were observed on the more hydrophilic
areas with the highest coordinating atoms (large number of neighbours) \cite{Yang2009}.
In this case, the well-defined topography of the HOPG substrate was ideal to
assess the effect of substrate roughness \cite{Yang2009}.
Thus, it was concluded that the substrate roughness appeared to favour the formation of NBs, in
contrast to smooth hydrophilic substrates,
such as dehydroxylated silicon oxide wafer substrates \cite{Yang2003}.
These experiments \cite{Yang2003,Yang2009} constituted an early indication for the significance of the contact-line pinning
in rough substrates for the stabilisation of surface NBs.

\begin{figure}
    \centering
    \includegraphics[width=\columnwidth]{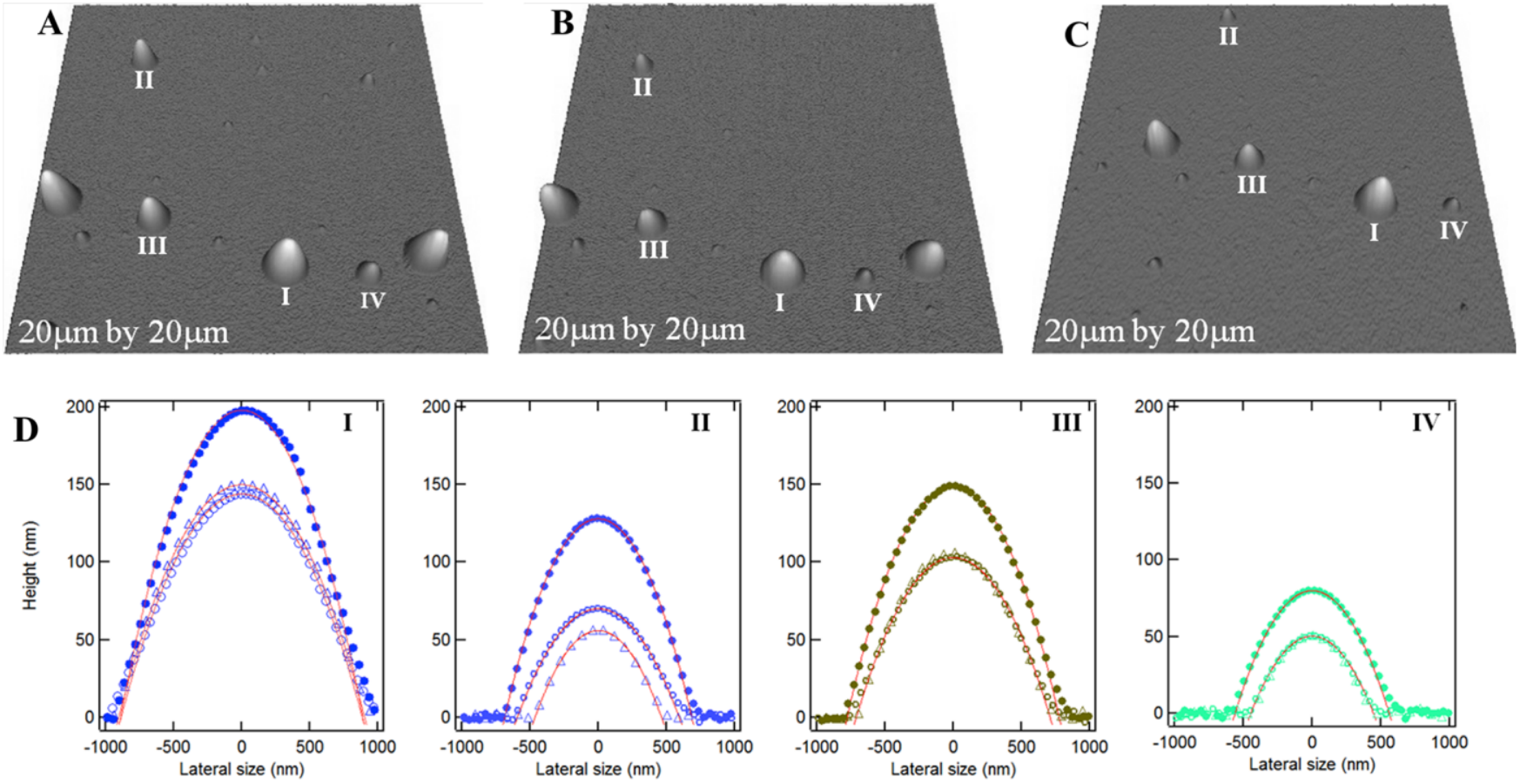}
    \caption{Contact line pinning. Morphology of surface NBs obtained from AFM and their fitting to spherical-cap shape. Reprinted with permission from Ref.~\cite{Zhang2013}. Copyright 2013 American Chemical Society.}
    \label{fig:figure06}
\end{figure}

Despite these earlier observations, the pinning assumption was established
by Zhang \textit{et al.}\ \cite{Zhang2013}, Weijs \textit{et al.}\ \cite{Weijs2013}, and Liu \textit{et al.}\  \cite{Liu2013b} .
By using AFM, Zhang \textit{et al.}\ suggested that the three-phase boundary of NBs were pinned during
their morphological evolution (see Fig.~\ref{fig:figure06}) \cite{Zhang2013}.
Moreover, the saturation levels of the dissolved gas affect the NB lifetime and
pinning of the contact line slows down the kinetics of both the
growth and shrinking processes.
Based on these observations, a bespoke 1D Epstein--Plesset model of gas diffusion was proposed,
which included the effect of pinning \cite{Zhang2013}.
However, the origin of the boundary of the pinning was still not completely understood
at the time.
A subsequent theoretical model tried to explain NBs stability by assuming a limited gas diffusion,
the cooperative effect of NB clusters and the pinned contact
line, which lead to a slower dissolution rate \cite{Weijs2013},
in agreement with recent theoretical arguments on the collective
dissolution of microbubbles \cite{Michelin2018}.
The model did not require fitting or uncontrolled assumptions to obtain the
lifetimes of NBs and predictions were in agreement
with the experimental findings \cite{Weijs2013}.
On the contrary, molecular-level simulations have indicated that NBs have shorter lifetimes
(\textit{i.e.} 100~ns) due to the limited diffusion stemming from the limited system sizes,
in contrast to the long lifetimes of experimental NBs \cite{Weijs2013,Che2017}.
While contact-line pinning is important for
the stability of single surface NBs, it also plays an important role in
suppressing the Ostwald ripening process between neighbouring NBs \cite{Dollet2016}, which could result in the growing of small NBs (large curvature) and the shrinking of large NBs (small curvature). This mechanism can
explain the different radii of curvature of NBs occurring in the
case of a population of neighbouring surface NBs.

The pinning force can be experimentally estimated  by pulling NBs with an AFM tip
and monitoring the mechanical response with Total Internal Reflection Fluorescence
Microscopy \cite{Tan2017a}. This force has been recently estimated of the order of
0.1~{\textmu}N, which is the force required to unpin the NB from its substrate \cite{Tan2017a}.
The force estimation
is limited by the stability of the neck pulled from the bubble and is enhanced by the
hydrophobicity of the tip. In particular,
the more hydrophobic the tip, the larger the pinning forces that can be measured \cite{Tan2017a}.
The measured pinning force is in agreement with previous theoretical work in the literature,
such as estimates from a lattice density functional theory \cite{Liu2014a}.

Evidence for the contact-line pinning also stems from
MD simulation of a CG model, which has demonstrated that surface NBs can be
stabilised in superheated or gas-oversaturated liquid
due to the substrate heterogeneity \cite{Liu2014b}.
In this case, the oversaturation refers to superheating for
pure liquids and gas oversaturation or superheating in the case of gas--liquid mixtures,
with both conditions having the same effect on NBs stability \cite{Liu2014b}.
Still, pinning is currently believed to be the main reason for the stability of surface NBs,
while hydrophobicity of the substrate
and oversaturation seem to play secondary roles \cite{Tan2018}.
Hence, in the model by Tan \textit{et al.}\ only the pinning of the contact line
is considered and is strictly required for the stability of NBs \cite{Tan2018}.
Still, hydrophobicity and oversaturation can enhance this stability, due to the
hydrophobic attraction between the substrate and the gas \cite{Tan2018},
in agreement with recent MD simulations \cite{Che2017}.
In fact, experiments have shown that NBs can also exist in open systems and undersaturated
environments \cite{Tan2018}.

Further work by numerical simulation has underlined the role of contact-line pinning in the stability of NBs.
Liu and Zhang proposed a mechanism for the three-phase contact-line pinning to obtain stable NBs,
which result from the intrinsic nanoscale physical roughness or chemical heterogeneity of the
substrate \cite{Liu2013b}.
By using classical DFT,
it has been shown that NBs are in thermodynamic metastable states \cite{Liu2013b},
in agreement with MD simulation \cite{Che2017}.
This theoretical assumption is consistent with
nucleation theory and can predict relationships between the contact angle and size of the NBs,
as well as the chemical potential \cite{Liu2013b}.
The critical nucleus can be stabilised due to pinning and its size can be estimated \cite{Liu2013b},
along the magnitude of the pinning force itself \cite{Liu2014a}.
In the case of stable NBs, the contact angles are independent of the substrate chemistry as its effects are
cancelled out by the pinning, in agreement with experimental
observations \cite{Liu2014a}. Moreover, a two-step process can explain the NB nucleation
based on the crevice model, \textit{i.e.} entrapped air pockets
in surface cavities that grow by diffusion \cite{Wang2017b}.
In the latter study by Wang \textit{et al.}, the authors have provided direct evidence of spontaneous NB formation as the
substrate is immersed in water. In this case, the size and
shape of the nanostructures play a role. For example, non-circular
nanopits lead to NBs with non-circular footprint, which shows
that strong pinning forces at the three-phase contact
line \cite{Wang2017b}. The effect of substrate hydrophobicity, where
NBs are formed by the ethanol--water exchange method has been
considered in the study by Zou \textit{et al.}\ by studying
an dodecyltrichlorosilane (DTS), an OTS, and an HOPG substrate
\cite{Zou2018}. The authors found that increased hydrophobicity
favours smaller contact angles (gas side), which is expected
given the hydrophobic character of the gas phase. Moreover,
nanoscopic and macroscopic contact angles match when the
substrate hydrophobicity increases \cite{Zou2018}.

Finally, the disjoining pressure originating from the van der Waals interactions
of the liquid and the gas with the substrate can affect the properties of NBs \cite{Svetovoy2016}.
In particular, it is believed that the disjoining pressure restricts the aspect ratio (lateral size/height)
of the NB and a maximal aspect ratio exists \cite{Svetovoy2016}.
The influence of the disjoining pressure on the NB shape
is minimal predicting a spherical-cap shape for individual NBs, in this way
confirming early assumptions from AFM experiments \cite{Svetovoy2016}.

\section{Perspectives}\label{sec:sec6-perspectives}
The main assumption for the unexpected high stability of surface NBs has been the contact-line
pinning caused by topological and chemical heterogeneities on the surfaces.
Still, AFM cannot distinguish between the
NBs and other objects and conclusions may be affected by this limitation.
Non-intrusive methods with both high spatial and temporal resolutions would be ideal.
Different simulation methods can provide details
on the mechanisms of NBs, such as MD, classical DFT, Monte Carlo, and others,
which may also provide further information on
various physical, chemical, mechanical, and thermodynamic properties of surface NBs.
Examples of various properties and parameters of interest may be the
distribution of charges, electric double layers, molecular structures, the presence of contaminants,
surface-active additives, surface energy (wettability), and others.
Moreover, there currently exists a limited selection of studied substrates in the literature, such as HOPG and graphene.
It would probably be worthwhile to conduct experiments on different substrates in the future.
Further experiments investigating
the interaction of NBs with external fields, such as electric and magnetic fields, temperature,
acoustic, and pressure waves could also be of interest for a number of applications in the
research area of surface NBs.

\section*{Acknowledgements}
This project has received funding from the European Union's Horizon 2020 research and innovation programme
under the Marie Sk{\l}odowska-Curie grant agreement No.\ 778104.
The work is also supported by the National Natural Science Foundation of China (Grant No.\ 51676137) and the Natural Science Foundation of Tianjin City (Grant No.\ 16JCYBJC41100).
This research is supported in part by
PLGrid Infrastructure.

%\singlespacing

\section*{References}

\bibliography{elsarticle-template}

\newpage

\newpage

\newpage

\newpage

\newpage

\newpage

\end{document}